\documentclass[aps,showpacs,epsfig,twocolumn]{revtex4}
\usepackage{epsfig}
\usepackage{appendix}

\newcommand{\be}{\begin{equation}}
\newcommand{\ee}{\end{equation}}
\newcommand{\bea}{\begin{eqnarray}}
\newcommand{\eea}{\end{eqnarray}}
\usepackage{color}
\begin{document}

\title{\bf\Large {Towards the theory of ferrimagnetism II}}

\author{Naoum Karchev }

\affiliation{Department of Physics, University of Sofia, 1126 Sofia, Bulgaria }

\begin{abstract}
The present paper is a sequel to the paper by Karchev (2008 J.Phys.:Condens.Matter {\bf 20} 325219).
A two-sublattice ferrimagnet, with spin-$s_1$ operators $\bf{S_{1i}}$ at the sublattice $A$ site and spin-$s_2$ operators $\bf{S_{2i}}$ at the sublattice $B$ site, is considered. Renormalized spin-wave theory, which accounts for the magnon-magnon interaction, and its extension are developed to describe the two ferrimagnetic phases $(0,T^*)$ and $(T^*,T_N)$ in the system, and to calculate the magnetization as a function of temperature.

The influence of the parameters in the theory on the characteristic temperatures $T_N$ and $T^*$ is studied. It is shown that, increasing the inter-sublattice exchange interaction, the ratio $T_N/T^*>1$ decreases approaching one, and above some critical value of the exchange constant there is only one phase $T_N\,=\,T^*$, and the magnetization-temperature curve has the typical Curie-Weiss profile. When the intra-exchange constant of sublattice with stronger intra-exchange interaction increases the $Ne\grave{e}l$ temperature increases while $T^*$ remains unchanged. Finally, when the magnetic order of the sublattice with smaller magnetic order decreases, $T^*$ decreases.  The theoretical predictions are utilize to interpret the experimentally measured magnetization-temperature curves.
\end{abstract}

\pacs{75.50.Bb, 75.30.Ds, 75.60.Ej, 75.50.-y}

\maketitle

\section {\bf Introduction}

The present paper is a sequel to the paper \cite{Karchev08b}.
A two-sublattice ferrimagnet, with spin-$s_1$ operators $\bf{S_{1i}}$ at the sublattice $A$ site and spin-$s_2$ operators $\bf{S_{2i}}$ at the sublattice $B$ site. The true magnons of a two-spin system are transversal fluctuations of the total magnetization which includes both the magnetization of the sublattice $A$ and $B$ spins.
The magnon excitation is a complicate mixture of the transversal
fluctuations of the sublattice $A$ and $B$ spins. As a result the magnons' fluctuations suppress,
in different way, the magnetic orders on the different sublattices and one obtains two phases.
At low temperature $(0,T^*)$ the magnetic orders of the $A$ and $B$ spins  contribute to the
magnetization of the system, while at the high temperature  $(T^*,T_N)$ the magnetization of the spins with a weaker intra-sublattice exchange is suppressed by magnon fluctuations, and only the spins with the stronger intra-sublattice exchange have non-zero spontaneous magnetization.

Renormalized spin-wave theory, which accounts for the magnon-magnon interaction, and its extension are developed to describe the two ferrimagnetic phases in the system and to calculate the magnetization as a function of temperature. It is impossible to require the theoretically calculated $N\acute{e}el$
temperature and magnetization-temperature curves to be in exact accordance with experimental results. The models are idealized, and they do not consider many important effects: phonon modes, several types of disorder, Coulomb interaction, etc. Because of this it is important to formulate
theoretical criteria for adequacy of the method of calculation. In my opinion the calculations should be in accordance with Mermin-Wagner theorem \cite{M-W}. It claims that in two dimensions there is not spontaneous magnetization at non-zero temperature. Hence, the critical temperature should be equal to zero. It is well known that the Monte Carlo method of calculation does not satisfy this criteria, and "weak z-coupling" 3D system is used to mimic a 2D layer. It is difficult within Dynamical Mean-Field Theory (DMFT) to make a difference between two dimensional and three dimensional systems. DMFT is a good approximation when the dimensionality goes to infinity. The present methods of calculation, being approximate, capture the basic physical features and satisfy the Mermin-Wagner theorem.

There is an important difference between $N\acute{e}el$ theory \cite{Neel} and the results in the present paper. $N\acute{e}el$'s calculations predict a temperature $T_N$ at which both the sublattice $A$ and $B$ magnetizations become equal to zero and $T^*$ is a temperature at which the magnetic moment has a maximum.

The influence of the parameters in the theory on the characteristic temperatures $T_N$ and $T^*$ is studied. It is shown that, increasing the inter-sublattice exchange interaction, the ratio $T_N/T^*>1$ decreases approaching one, and above some critical value of the exchange constant there is only one phase $T_N\,=\,T^*$, and the magnetization-temperature curve has the typical Curie-Weiss profile. When the intra-exchange constant of the sublattice with stronger intra-exchange interaction increases the $Ne\grave{e}l$ temperature increases while $T^*$ remains unchanged. Finally, when the magnetic order of the sublattice with smaller magnetic order decreases, $T^*$ decreases.

To compare the theoretical results  and the experimental magnetization-temperature curves one has, first of all, to interpret adequately the measurements. The magnetic moments in some materials are close to "spin only" value $2\mu_B S$ and the sublattice spins $s_1$ and $s_2$ can be obtained from the experimental curves. As an example I consider the sulpho-spinel $MnCr_2S_{4-x}Se_{x}$ \cite{HBMM3}. On the other hand there are ferrimagnets with strong spin-orbital interaction. It is convenient, in that case, to consider  $jj$ coupling with $\textbf{J}^A=\textbf{L}^A+\textbf{S}^A$ and $\textbf{J}^B=\textbf{L}^B+\textbf{S}^B$. As an example I consider the vanadium spinel $MnV_2O_4$\cite{vanadium1,vanadium2,vanadium3,vanadium4}.

The paper is organized as follows. In Sec. II the model is presented and a renormalized spin-wave theory is worked out to calculate the magnetization-temperature curves for different parameters of the model.
The influence of the theory parameters on the $N\acute{e}el$ and $T^*$ temperatures is studied in Sec. III. I consider three cases: i) when the inter-sublattice exchange constant increases and all the other parameters are fixed, ii) one of the intra-sublattice parameters is changed and iii) when one of the spins decreases.  Applications and analyzes of experimental magnetization-temperature curves are given in Sec. IV.
A summary in Sec. V concludes the paper.

\section {\bf Spin-wave theory}
\subsection{\bf Renormalized spin-wave (RSW) theory}

The Hamiltonian of the system is
\bea \label{rsw1}
 H & = & - J_1\sum\limits_{\ll ij \gg _A } {{\bf S}_{1i}
\cdot {\bf S}_{1j}}\,-\,J_2\sum\limits_{\ll ij \gg _B } {{\bf S}_{2i}
\cdot {\bf S}_{2j}}\nonumber \\
& & +\,J \sum\limits_{\langle ij \rangle} {{\bf S}_{1i}
\cdot {\bf S}_{2j}}
  \eea where the sums are over all sites of a three-dimensional cubic lattice:
$\langle i,j\rangle$ denotes the sum over the nearest neighbors, $\ll i,j \gg _A$ denotes the sum over the sites of the A sublattice, $\ll i,j \gg _B$ denotes the sum over the sites of the B sublattice.
The first two terms  describe the ferromagnetic Heisenberg intra-sublattice
exchange $J_1>0, J_2>0$, while the third term describes the inter-sublattice exchange which is antiferromagnetic $J>0$.
To study a theory with the Hamiltonian Eq.(\ref{rsw1}) it is convenient to introduce Holstein-Primakoff representation for the spin
operators
\bea\label{rsw2} & &
S_{1j}^+ = S^1_{1j} + i S^2_{1j}=\sqrt {2s_1-a^+_ja_j}\,\,\,\,a_j \nonumber \\
& & S_{1j}^- = S^1_{1j} - i S^2_{1j}=a^+_j\,\,\sqrt {2s_1-a^+_ja_j}
\\ & & S^3_{1j} = s_1 - a^+_ja_j \nonumber \eea
when the sites $j$ are from sublattice $A$ and
\bea\label{rsw3} & &
S_{2j}^+ = S^1_{2j} + i S^2_{2j}=-b^+_j\,\,\sqrt {2s_2-b^+_jb_j}\nonumber \\
& & S_{2j}^- = S^1_{2j} - i S^2_{2j}=-\sqrt {2s_2-b^+_jb_j}\,\,\,\,b_j
\\ & & S^3_{2j} = -s_2 + b^+_jb_j \nonumber \eea
when the sites $j$ are from sublattice $B$. The operators $a^+_j,\,a_j$ and  $b^+_j,\,b_j$ satisfy the Bose commutation relations.
In terms of the Bose operators and keeping only the quadratic and quartic terms, the effective Hamiltonian
Eq.(\ref{rsw1}) adopts the form
\be\label{rsw4}H=H_2+H_4\ee where
\bea\label{rsw5}
 H_2 & = & s_1 J_1\sum\limits_{\ll ij \gg _A }\left( a^+_i a_i\,+\,a^+_j a_j\,-\,a^+_j a_i\,-\,a^+_i a_j\right) \nonumber \\
  & + & s_2 J_2\sum\limits_{\ll ij \gg _B }\left( b^+_i b_i\,+\,b^+_j b_j\,-\,b^+_j b_i\,-\,b^+_i b_j\right) \\
 & + & J \sum\limits_{\langle ij \rangle}\left[s_1 b^+_j b_j + s_2 a^+_i a_i -
 \sqrt{s_1 s_2}\left( a^+_i b^+_j+a_i b_j \right)\right] \nonumber
 \eea
 \bea\label{rsw6}
 H_4 & = & \frac 14 J_1 \sum\limits_{\ll ij \gg _A }\left[a^+_i a^+_j( a_i-a_j)^2 + (a^+_i- a^+_j)^2  a_i a_j\right] \nonumber \\
 & + & \frac 14 J_2 \sum\limits_{\ll ij \gg _B }\left[b^+_i b^+_j( b_i-b_j)^2 + (b^+_i- b^+_j)^2  b_i b_j\right]\nonumber \\
 & + & \frac 14 J \sum\limits_{\langle ij \rangle}\left[
 \sqrt{\frac {s_1}{s_2}}\left( a_i b^+_j b_j b_j+a^+_i b^+_j b^+_j b_j \right)\right. \\
& + &\left. \sqrt{\frac {s_2}{s_1}}\left(a^+_i a_i a_i b_j+a^+_i a^+_i a_i b^+_j \right) - 4 a^+_i a_i b^+_j b_j \right] \nonumber
 \eea
and the terms without operators are dropped.

The next step is to represent the Hamiltonian in the Hartree-Fock  approximation
\be\label{rsw7}H\approx H_{HF}=H_{cl}+H_q\ee where
\bea\label{rsw8} H_{cl} & = & 12 N J_1 s_1^2 (u_1-1)^2+ 12 N J_2 s_2^2 (u_2-1)^2 \nonumber \\
& + & 6 N J s_1 s_2 (u-1)^2,
\eea
\bea\label{rsw9}
 H_2 & = & s_1 J_1 u_1\sum\limits_{\ll ij \gg _A }\left( a^+_i a_i\,+\,a^+_j a_j\,-\,a^+_j a_i\,-\,a^+_i a_j\right) \nonumber \\
  & + & s_2 J_2 u_2\sum\limits_{\ll ij \gg _B }\left( b^+_i b_i\,+\,b^+_j b_j\,-\,b^+_j b_i\,-\,b^+_i b_j\right) \\
 & + & J u\sum\limits_{\langle ij \rangle}\left[s_1 b^+_j b_j + s_2 a^+_i a_i -
 \sqrt{s_1 s_2}\left( a^+_i b^+_j+a_i b_j \right)\right] \nonumber
 \eea
and $N=N_A=N_B$ is the number of sites on a sublattice. Equation (\ref{rsw9}) shows that the Hartree-Fock parameters $u_1,\,u_2$ and $u$ renormalize
the intra-exchange constants $J_1,\,J_2$ and the inter-exchange constant $J$, respectively.

It is convenient to rewrite the Hamiltonian in momentum space representation
\be\label{rsw10}
H_q = \sum\limits_{k\in B_r}\left [\varepsilon^a_k\,a_k^+a_k\,+\,\varepsilon^b_k\,b_k^+b_k\,-
 \,\gamma_k \left (a_k^+b_k^+ + b_k a_k \right )\,\right ],
\ee
where the wave vector $k$ runs over the reduced  first Brillouin zone $B_r$ of a
cubic lattice. The dispersions are given by equalities
\bea\label{rsw11}
\varepsilon^a_k & = & 4s_1\,J_1\,u_1\varepsilon_k
\,+\,6s_2\,J u \nonumber\\
\varepsilon^b_k & = & 4s_2\,J_2\,u_2 \varepsilon_k \,+\,6s_1\,J\,u \\
\gamma_k & = & 2J\,u\,\sqrt{s_1\,s_2}\,\left(\cos k_x +\cos k_y + \cos k_z \right)\nonumber\eea with
\bea\label{rsw12}\varepsilon_k & = & 6-\cos(k_x+k_y)-\cos(k_x-k_y) - \cos (k_x+k_z) \nonumber\\
 & - & \cos(k_x-k_z)- \cos(k_y+k_z) - \cos(k_y-k_z).\eea To diagonalize the Hamiltonian one introduces new Bose fields
$\alpha_k,\,\alpha_k^+,\,\beta_k,\,\beta_k^+$ by means of the
transformation
\bea \label{rsw12a} & &
a_k\,=u_k\,\alpha_k\,+\,v_k\,\beta^+_k\qquad
a_k^+\,=u_k\,\alpha_k^+\,+\,v_k\,\beta_k
\nonumber \\
\\
& & b_k\,=\,u_k\,\beta_k\,+\,v_k\,\alpha^+_k\qquad
b_k^+\,=\,u_k\,\beta_k^+\,+\,v_k\,\alpha_k
\nonumber \eea where the coefficients of the transformation $u_k$ and $v_k$ are real function of the wave vector $k$
\bea \label{ferri11} &
&u_k\,=\,\sqrt{\frac 12\,\left (\frac
{\varepsilon^a_k+\varepsilon^b_k}{\sqrt{(\varepsilon^a_k+\varepsilon^b_k)^2-4\gamma^2_k}}\,+\,1\right
)}\nonumber \\
\\
& & v_k\,=\,sign (\gamma_k)\,\sqrt{\frac 12\,\left (\frac
{\varepsilon^a_k+\varepsilon^b_k}{\sqrt{(\varepsilon^a_k+\varepsilon^b_k)^2-4\gamma^2_k}}\,-\,1\right
)}\nonumber \eea
The transformed Hamiltonian adopts the form \be
\label{ferri12} H_q = \sum\limits_{k\in B_r}\left
(E^{\alpha}_k\,\alpha_k^+\alpha_k\,+\,E^{\beta}_k\,\beta_k^+\beta_k\,+\,E^0_k\right),
\ee
with new dispersions \bea  \label{ferri13} & & E^{\alpha}_k\,=\,\frac
12\,\left [
\sqrt{(\varepsilon^a_k\,+\,\varepsilon^b_k)^2\,-\,4\gamma^2_k}\,-\,\varepsilon^b_k\,+\,\varepsilon^a_k\right] \nonumber \\
\\
& & E^{\beta}_k\,=\,\frac
12\,\left [
\sqrt{(\varepsilon^a_k\,+\,\varepsilon^b_k)^2\,-\,4\gamma^2_k}\,+\,\varepsilon^b_k\,-\,\varepsilon^a_k\right]
\nonumber \eea and vacuum energy
\be\label{ferri14}
 E^{0}_k\,=\,\frac
12\,\left [
\sqrt{(\varepsilon^a_k\,+\,\varepsilon^b_k)^2\,-\,4\gamma^2_k}\,-\,\varepsilon^b_k\,-\,\varepsilon^a_k\right]\ee

For positive values of the Hartree-Fock parameters and all values of $k\in B_r$,\,
the dispersions are nonnegative $ E^{\alpha}_k\geq 0,\, E^{\beta}_k \geq 0$. For definiteness I choose $s_1>s_2$.
With these parameters, the $\alpha_k$ boson is the long-range \textbf{(magnon)} excitation in the two-spin system with $E^{\alpha}_k\propto\rho k^2$, near the zero wavevector, while the $\beta_k$ boson is a gapped excitation.

To obtain the system of equations for the Hartree-Fock parameters we consider
the free energy of a system with Hamiltonian $H_{HF}$ equations (\ref{rsw8}) and (\ref{ferri12})
\bea\label{rsw12}
\mathcal{F} & = & 12J_1 s_1^2 (u_1-1)^2+ 12J_2 s_2^2 (u_2-1)^2  \nonumber \\
& + & 6J s_1 s_2 (u-1)^2 + \frac 1N \sum\limits_{k\in B_r}E^{0}_k \\
& + & \frac {1}{\beta N} \sum\limits_{k\in B_r}\left[ \ln\left(1-e^{-\beta E^{\alpha}_k}\right)\,+\,\ln\left(1-e^{-\beta E^{\beta}_k}\right)\right].\nonumber\eea
where $\beta\,=\,1/T$\,\, is the inverse temperature. Then the three equations
\be\label{rsw13}\partial\mathcal{F}/\partial u_1=0,\quad \partial\mathcal{F}/\partial u_2=0,\quad\partial\mathcal{F}/\partial u=0\ee adopt the form
 (see the appendix)
\bea\label{rsw14} u_1 & = & 1-\frac {1}{6s_1} \frac 1N \sum\limits_{k\in B_r} \varepsilon_k \left[u_k^2 \,n_k^{\alpha}\, +\, v_k^2\, n_k^{\beta}\, +\, v_k^2\right]\nonumber \\
u_2 & = & 1-\frac {1}{6s_2} \frac 1N \sum\limits_{k\in B_r} \varepsilon_k \left[v_k^2 \,n_k^{\alpha}\, +\, u_k^2\, n_k^{\beta}\, +\, v_k^2\right]\nonumber \\
u & = & 1-\frac 1N \sum\limits_{k\in B_r} \left[\frac {1}{2s_1}\left(u_k^2 \,n_k^{\alpha}\, +\, v_k^2\, n_k^{\beta}\, +\, v_k^2\right)\right. \\
& + & \left. \frac {1}{2s_2} \left(v_k^2 \,n_k^{\alpha}\, +\, u_k^2\, n_k^{\beta}\, +\, v_k^2\right)\right. \nonumber \\
& - & \left.\frac 23 J u\left(1+n_k^{\alpha}+n_k^{\beta}\right) \frac {\left(\cos k_x +\cos k_y + \cos k_z \right)^2}{\sqrt{(\varepsilon^a_k\,+\,\varepsilon^b_k)^2\,-\,4\gamma^2_k}}
\right]\nonumber
\eea
where $n_k^{\alpha}$ and $n_k^{\beta}$ are the Bose functions of $\alpha$ and $\beta$ excitations. The Hartree-Fock parameters, the solution of the system of equations (\ref{rsw14}), are positive functions of $T/J$, $u_1(T/J)>0,\,u_2(T/J)>0$ and $u(T/J)>0$. Utilizing these functions, one can calculate the spontaneous magnetization of the system, which is a sum of the spontaneous magnetization on the two sublattices $M\,=\,M^A\,+\,M^B$, where
\bea \label{rsw15}
M^A & = & <S^3_{1j}> \,\,\, j\,\, is\,\, from\,\, sublattice\,\, A \nonumber \\
\\
M^B & = & <S^3_{2j}> \,\,\, j\,\, is\,\, from\,\, sublattice\,\, B \nonumber \eea
In terms of the Bose functions of the $\alpha$ and $\beta$ excitations they adopt the form
\bea\label{rsw16}
M^A & = & s_1\,-\,\frac 1N \sum\limits_{k\in B_r} \left[u_k^2 \,n_k^{\alpha}\, +\, v_k^2\, n_k^{\beta}\, +\, v_k^2\right] \\
M^B & = & - \,s_2\,+\,\frac 1N \sum\limits_{k\in B_r} \left[v_k^2 \,n_k^{\alpha}\, +\, u_k^2\, n_k^{\beta}\, +\, v_k^2\right]\nonumber \eea

The magnon excitation-$\alpha_k$ in the effective theory equation (\ref{ferri12})- is a complicated mixture of the transversal
fluctuations of the $A$ and $B$ spins. As a result the magnons' fluctuations suppress in a different way the magnetization on sublattices $A$ and $B$. Quantitatively this depends on the coefficients $u_k$ and $v_k$ in equations (\ref{rsw16}). At characteristic temperature $T^*$  spontaneous magnetization on sublattice $B$ becomes equal to zero, while spontaneous magnetization on sublattice $B$ is still nonzero. Above $T^*$ the system of equations (\ref{rsw14}) has no solution and one has to modify the spin-wave theory. The magnetization depends on the
dimensionless temperature $T/J$ and dimensionless parameters
$s_1,\,s_2,\,J_1/J$ and $J_2/J$. For parameters $s_1=1.5,\,s_2=1,\,J_1/J=0.94$ and $J_2/J=0.01$ the
functions $M^A(T/J)$ and $M^B(T/J)$ are depicted in
figure 1.  The upper (blue) line is the sublattice $A$ magnetization, the bottom (red) line is the sublattice $B$ magnetization.
\begin{figure}[!ht]
\epsfxsize=8.cm 
\epsfbox{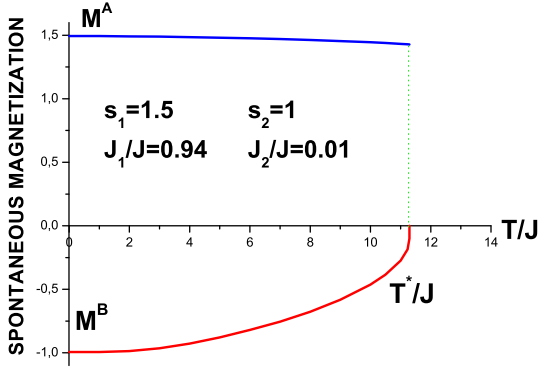} \caption{(color online)\,The spontaneous
magnetization $M^A$-upper (blue) line and $M^B$-bottom (red) line as
a function of $T/J$ for parameters\,
$s_1\,=\,1.5,\,s_2\,=\,1,\,J_1/J\,=\,0.94$ and $J_2/J\,=\,0.01$.
$T^*$ is the temperature at which sublattice $B$ magnetization
becomes equal to zero}\label{fig1}
\end{figure}

\subsection{\bf Modified RSW theory}

Once suppressed, the sublattice $B$ magnetization cannot be restored increasing the temperature above T*. To formulate this mathematically
we modify the spin-wave theory using the idea of a description of the paramagnetic phase of 2D ferromagnets ($T>0$) by means of modified spin-wave
theory \cite{Takahashi1,Takahashi2} and its generalization \cite{Karchev08b}. We consider a two-sublattice system and, to enforce the magnetization on the two sublattices to be equal to zero in paramagnetic phase,
we introduce two parameters $\lambda_A$ and $\lambda_B$ \cite{Karchev08b}. The new Hamiltonian is obtained from the old one equation (\ref{rsw1}) by adding two new terms:
\be
\label{ferri26} \hat{H}\,=\,H\,-\,\sum\limits_{i\in A}
\lambda_1 S^3_{1i}\,+\,\sum\limits_{i\in B} \lambda_2 S^3_{2i} \ee
In momentum space
the new Hamiltonian adopts the form \be \label{ferri27}
 \hat{H} = \sum\limits_{k\in B_r}\left [\hat{\varepsilon}^a_k\,a_k^+a_k\,+\,\hat{\varepsilon}^b_k\,b_k^+b_k\,-
 \,\gamma_k\,(b_k a_k+b_k^+ a_k^+)\right] \ee
where the new dispersions are
 \be \label{ferri28}
\hat{\varepsilon}^a_k\,=\varepsilon^a_k\,+\,\lambda_1, \qquad
\hat{\varepsilon}^b_k\,=\varepsilon^b_k\,+\,\lambda_2.\ee
Utilizing the same
transformation equations (\ref{rsw12a}) with parameters
\bea \label{ferri29} &
&\hat{u}_k\,=\,\sqrt{\frac 12\,\left (\frac
{\hat{\varepsilon}^a_k+\hat{\varepsilon}^b_k}{\sqrt{(\hat{\varepsilon}^a_k+\hat{\varepsilon}^b_k)^2-4\gamma^2_k}}\,+\,1\right
)}\nonumber \\
\\
& &\hat{v}_k\,=\,sign (\gamma_k)\,\sqrt{\frac 12\,\left (\frac
{\hat{\varepsilon}^a_k+\hat{\varepsilon}^b_k}{\sqrt{(\hat{\varepsilon}^a_k+\hat{\varepsilon}^b_k)^2-4\gamma^2_k}}\,-\,1\right
)}\nonumber \eea
one obtains the Hamiltonian in diagonal forma
\be
\label{ferri30} \hat{H} = \sum\limits_{k\in B_r}\left
(\hat{E}^{\alpha}_k\,\alpha_k^+\alpha_k\,+\,\hat{E}^{\beta}_k\,\beta_k^+\beta_k+\hat{E}^0_k\right),
\ee where
\bea\label{ferri31} & & \hat{E}^{\alpha}_k\,=\,\frac
12\,\left [
\sqrt{(\hat{\varepsilon}^a_k\,+\,\hat{\varepsilon}^b_k)^2\,-\,4\gamma^2_k}\,-\,\hat{\varepsilon}^b_k\,+\,\hat{\varepsilon}^a_k\right] \nonumber \\
& & \hat{E}^{\beta}_k\,=\,\frac
12\,\left [
\sqrt{(\hat{\varepsilon}^a_k\,+\,\hat{\varepsilon}^b_k)^2\,-\,4\gamma^2_k}\,+\,\hat{\varepsilon}^b_k\,-\,\hat{\varepsilon}^a_k\right]\\
& & \hat{E}^{0}_k\,=\,\frac
12\,\left [
\sqrt{(\hat{\varepsilon}^a_k\,+\,\hat{\varepsilon}^b_k)^2\,-\,4\gamma^2_k}\,-\,\hat{\varepsilon}^b_k\,-\,\hat{\varepsilon}^a_k\right]\nonumber\eea

It is convenient to represent the parameters
$\lambda_1$ and $\lambda_2$ in the form \be \label{ferri32}
\lambda_1\,=\,6 J u s_2 (\mu_1\,-\,1),\quad
\lambda_2\,=\,6 J u s_1 (\mu_2\,-\,1). \ee In terms of the new
parameters $\mu_1$ and $\mu_2$ the dispersions
$\hat{\varepsilon}^a_k$ and $\hat{\varepsilon}^b_k$ adopt the form
\bea \label{ferri33} & & \hat{\varepsilon}^a_k\,=\,4 s_1
J_1\,u_1\,\varepsilon_k\,+\,6\,s_2\,J\,u\,\mu_1
\nonumber \\
\\
& &
\hat{\varepsilon}^b_k\,=\,4s_2\,J_2\,u_2\,\varepsilon_k\,+\,6\,s_1\,J\,u\,\mu_2
\nonumber \eea
They are positive ($\hat{\varepsilon}^a_k>0$, $\hat{\varepsilon}^b_k>0$) for all values of the wavevector $k$, if the parameters  $\mu_1$ and $\mu_2$ are positive ($\mu_1>0,\,\mu_2>0$).
The dispersions Eq.(\ref{ferri31}) are well defined if square-roots in equations
(\ref{ferri31}) are well defined. This is true if \be\label{ferri34} \mu_1
\mu_2\geq1.\ee The $\beta_k$ excitation is gapped ($E^{\beta}_k>0 $)
for all values of parameters $\mu_1$ and $\mu_2$ which satisfy equation
(\ref{ferri34}). The $\alpha$ excitation is gapped if $\mu_1
\mu_2>1$, but in the particular case \be \label{ferri35} \mu_1
\mu_2=1\ee $\hat{E}^{\alpha}_0=0$, and near the zero wavevector
\be \label{ferri35b}
\hat{E}^{\alpha}_k\approx \hat{\rho} k^2\ee with
spin-stiffness constant
\be \label{ferri35c} \hat{\rho}\,=\,\frac {8(s_2^2 J_2 u_2\mu_1
\,+\,s_1^2 J_1 u_1 \mu_2)\,+\,2s_1s_2J u}{(s_1 \mu_2-s_2\mu_1)}\ee In
the particular case equation (\ref{ferri35}) $\alpha_k$ boson is the
long-range excitation (magnon) in the system.

We introduced the parameters $\lambda_1$ and $\lambda_2$ ($\mu_1, \mu_2$) to enforce the sublattice $A$ and $B$ spontaneous magnetizations
to be equal to zero in the paramagnetic phase. We find out the parameters $\mu_1$ and $\mu_2$, as well as the Hartree-Fock parameters, as functions of temperature, solving the system of five equations, equations (\ref{rsw14}) and the equations $M^A=M^B=0$, where the  spontaneous magnetizations have the same representation as equations (\ref{rsw16}) but with coefficients $\hat{u}_k,\,\,
\hat{v}_k$, and dispersions $\hat{E}^{\alpha}_k,\,\, \hat{E}^{\beta}_k$ in the expressions for the Bose functions.
The numerical calculations show that for high enough temperature
$\mu_1\mu_2>1$. When the temperature decreases the product $\mu_1\mu_2$ decreases, remaining larger than one. The temperature at which the product becomes equal to one ($\mu_1\mu_2=1$) is the $N\acute{e}el$ temperature. Below $T_N$, the spectrum contains long-range (magnon) excitations, thereupon $\mu_1\mu_2=1$. It is convenient to represent the parameters in the following way:
\be\label{MnV16}\mu_1=\mu, \quad\quad \mu_2=1/\mu.\ee

In the ordered phase magnon excitations are the origin of the suppression of the magnetization. Near the zero temperature their contribution is
small and at zero temperature sublattice $A$ and $B$ spontaneous magnetization reach their saturation. On increasing the temperature magnon fluctuations suppress the sublattice $A$ magnetization and sublattice $B$ magnetization in different ways. At $T^*$ the sublattice $B$ spontaneous magnetization becomes equal to zero.
Increasing the temperature above $T^*$, the sublattice $B$ magnetization should be zero. This is why we impose the condition
$M^B(T)=0$ if $T>T^{*}$. For temperatures above $T^*$, the parameter $\mu$ and the Hartree-Fock parameters are solution of a system of four equations, equations (\ref{rsw14}) and the equation $M^B=0$. The Hartree-Fock parameters, as a functions of temperature $T/J$, are depicted in figure 2 for parameters $s_1=1.5,\,s_2=1,\,J_1/J=0.94$ and $J_2/J=0.01$.  The vertical dotted (green) line corresponds to $T^*/J$.
\begin{figure}[!ht]
\epsfxsize=7.5cm \hskip -0.9cm \epsfbox{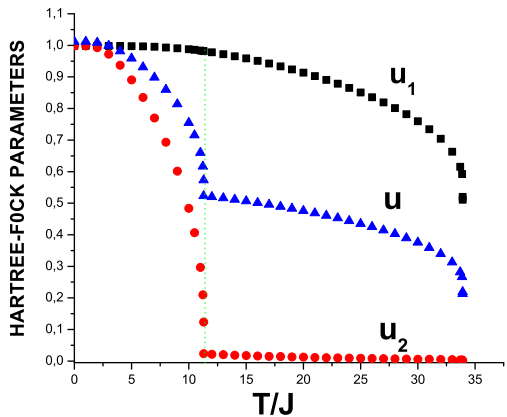}
\caption{(color online)\,Hartree-Fock parameters $u_1$,\,$u_2$ and
$u$ as a function of $T/J$ for
$s_1\,=\,1.5,\,s_2\,=\,1,\,J_1/J\,=\,0.94$ and $J_2/J\,=\,0.01$. The
vertical dotted (green) line corresponds to $T^*/J$  }\label{fig2}
\end{figure}

The function $\mu(T/J)$ is depicted in figure 3 for the same parameters.
\begin{figure}[!ht]
\epsfxsize=7.5cm \hskip -0.9cm \epsfbox{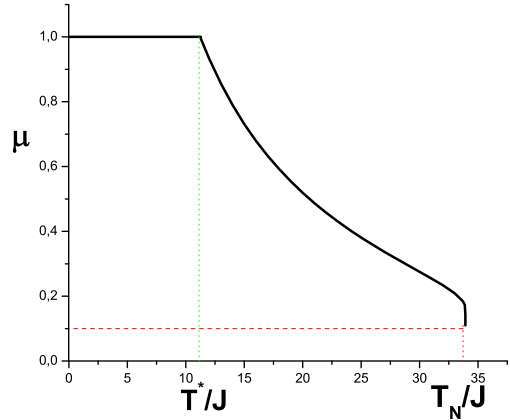} \caption{(color
online)\,$\mu(T/J)$ for parameters
$s_1\,=\,1.5,\,s_2\,=\,1,\,J_1/J\,=\,0.94$ and $J_2/J\,=\,0.01$. The
vertical dotted (green) line corresponds to $T^*/J$, while (red)
dashed lines to $T_N/J$ and $\mu(T_N/J)$.}\label{fig2}
\end{figure}

We utilize the
obtained function $\mu(T)$, $u_1(T)$, $u_2(T)$, $u(T)$ to calculate the spontaneous magnetization as a function of the temperature. Above $T^*$, the magnetization of the system is equal to the sublattice $A$ magnetization. For the same parameters as above the
functions $M^A(T/J)$ and $M^B(T/J)$ are depicted in figure 4a.
The upper (blue) line is the sublattice $A$ magnetization, the bottom (red) line is the sublattice $B$ magnetization. The total magnetization $M\,=\,M^A\,+\,M^B$ is depicted in figure 4b.
\begin{figure}[!ht]
\epsfxsize=7.8cm \hskip -0.9cm \epsfbox{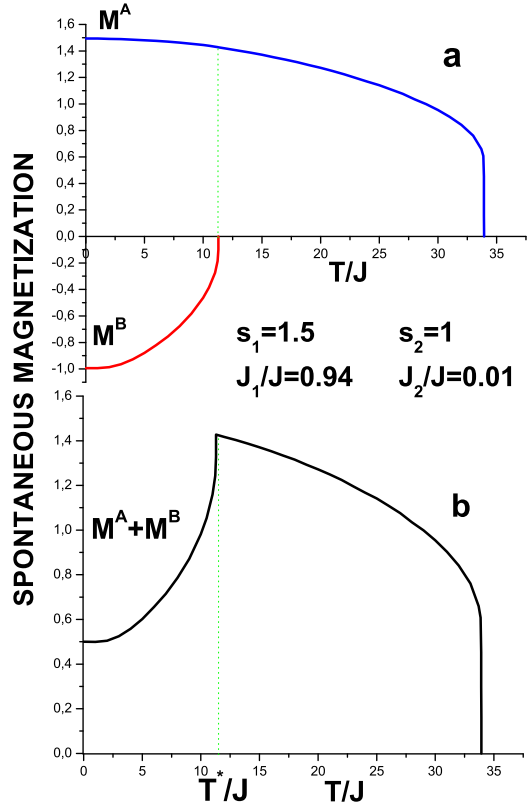} \caption{(color
online)\,a) The sublattice $A$ spontaneous magnetization $M^A$-upper
(blue) line and sublattice $B$ spontaneous magnetization
$M^B$-bottom (red) line as a function of $T/J$ for parameters\,
$s_1\,=\,1.5,\,s_2\,=\,1,\,J_1/J\,=\,0.94$ and
$J_2/J\,=\,0.01$.\,\\b) The total spontaneous magnetization
$M^A\,+\,M^B$.  $T^*/J$- vertical dotted (green) line}\label{fig4}
\end{figure}

\section{\bf  $T_N$ and $T^*$ dependence on model's parameters}

The existence of two ferromagnetic phases $(0,T^*)$ and $(T^*,T_N)$ is a generic feature of two spin systems. The characteristic temperatures $T_N$ and $T^*$ strongly  depend on the parameters of the model. Intuitively, it is clear that, if the inter-exchange is much stronger than intra-exchanges, the ferromagnetic order sets in simultaneously on both sublattices. This is not true, if inter-exchange is not so strong. To demonstrate this I study a system with sublattice $A$ spin $s_1=1.5$, and sublattice $B$ spin $s_2=1$. For parameters $J_1/J=0.5$ and $J_2/J=0.005$ the magnetization-temperature curve is depicted in FIG.5 curve "c". The ratio of the characteristic temperatures equals $T_N/T^*\,=\,1.722$. Increasing the inter-exchange coupling, $J_1/J=0.3$,\,$J_2/J=0.003$ (curve "b"), the ratio decreases $T_N/T^*\,=\,1.229$, and above some critical value of the inter-exchange constant $J_1/J=0.05$,\,$J_2/J=0.0005$ $N\acute{e}el$'s temperature becomes equal to $T^*$. There is only one ferromagnetic phase, and magnetization-temperature curve "a" is a typical Curie-Weiss curve. Despite this the system does not describe ferromagnet, because the spin wave excitations are superposition of the sublattice $A$ and $B$ spin excitations.
\begin{figure}[!ht]
\epsfxsize=9.5cm \hskip -0.8cm \epsfbox{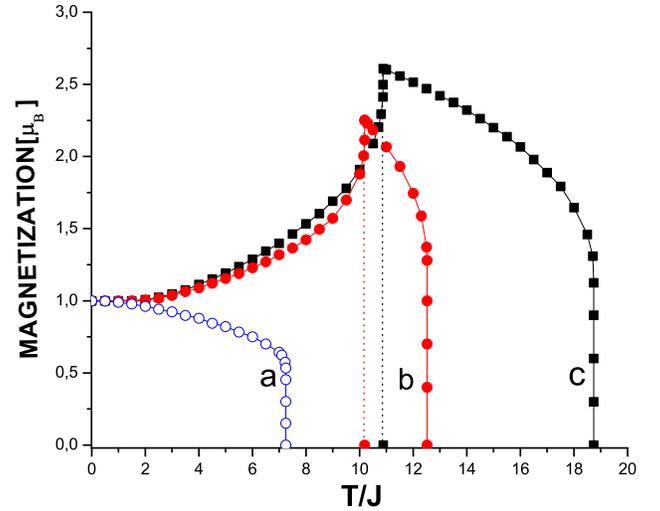} \caption{(color
online)\, The magnetization $2\,M^A\,+\,2\,M^B$ as a function of
$T/J$ for $s_1=1.5$ and $s_2\,=\,1$, curve
\textbf{a}:$J_1/J\,=\,0.05$,\,$J_2/J\,=\,0.0005$, curve
\textbf{b}:$J_1/J\,=\,0.3$,\,$J_2/J\,=\,0.003$, curve
\textbf{c}:$J_1/J\,=\,0.5$,\,$J_2/J\,=\,0.005$.}\label{fig5}
\end{figure}

Next, I consider a system with sublattice $A$ spin $s_1\,=\,1.5$, and sublattice $B$ spin $s_2\,=1\,$. The ratio of sublattice $B$ exchange constant $J_2$ and inter-exchange constant $J$ is fixed $j_2\,=\,J_2/J\,=\,0.01$, while the ratio $j_1\,=\,J_1/J$ varies.  When the sublattice $A$ exchange constant $J_1$ increases $j_1=J_1/J\,=\,0.64,\,0.84,\,0.94$, the magnetization-temperature curve at temperatures below $T^*$ does not change. There is no visible difference between $T^*$ temperatures for the three values of the parameter $J_1/J$. The difference appears when the temperature is above $T^*$. Increasing sublattice $A$ exchange constat increases the $N\acute{e}el$ temperature. The three curves are depicted in figure 6.
\begin{figure}[!ht]
\epsfxsize=9.2cm \hskip -0.3cm \epsfbox{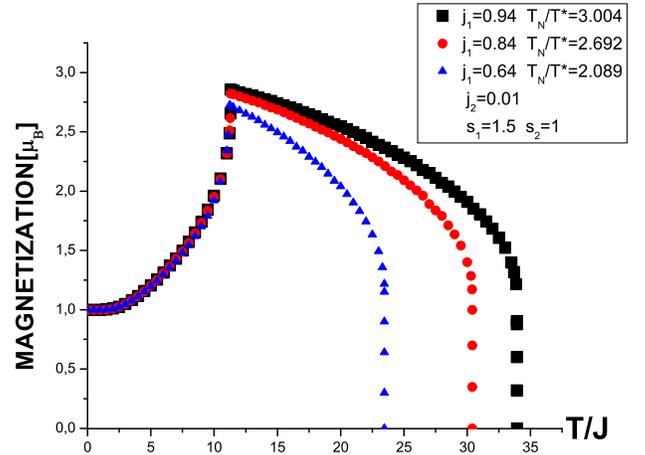} \caption{(color
online)\, The magnetization $2\,M^A\,+\,2\,M^B$ as a function of
$T/J$ for $s_1=1.5$,\,\,$s_2\,=\,1$,\,\,$j_2\,=\,J_2/J\,=\,0.01$ and
three values of the parameter $j_1\,=\,J_1/J$; $j_1\,=\,0.94$
(black) squares,\, $j_1\,=\,0.84$ (red) circles,\, $j_1\,=\,0.64$
(blue) triangles}\label{fig6}
\end{figure}
\begin{figure}[!ht]
\epsfxsize=9.cm \hskip -0.3cm \epsfbox{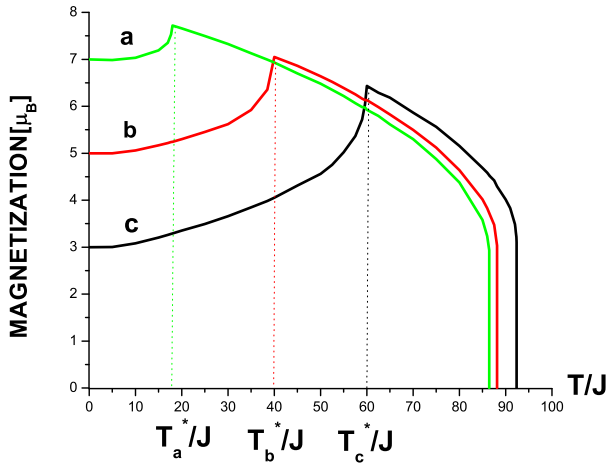} \caption{(color
online)\, The magnetization $2M^A+2M^B$ as function of $T/J$ for
$J_1/J=0.4$, $J_2/J\,=\,0.004$,\, $s_1=4$, and $s_2=0.5$-curve
\textbf{a}\,(green), $s_2=1.5$-curve \textbf{b}\,(red),
$s_2=2.5$-curve \textbf{c}\,(black)}\label{fig7}
\end{figure}

Finally, I consider three systems with equal exchange constants $J_1/J\,=\,0.4$, $J_2/J=0.004$ and sublattice $A$ spin $s_1=4$, but with three different sublattice $B$ spins (figure 7). The calculations show that decreasing the sublattice $B$ spin decreases $T^*$ temperature, increases the maximum of magnetization at $T^*$ and zero temperature magnetization.

\section{\bf Theory and experiment}
\subsection{\bf Sulpho-spinel $MnCr_2S_{4-x}Se_x$}

The sulpho-spinel $MnCr_2S_{4-x}Se_x$ has been investigated by measurements of the magnetization at $15.3kOe$ as a function of temperature (figure 94 in \cite{HBMM3}). The maximum in the magnetization versus temperature curve, which is typical of $MnCr_2S_4$ ($x=0$),
increases when $x$ increase, and disappears at $x=0.5$. The $N\acute{e}el$ temperature decreases from $74K$ at $x=0$ to $56K$ at $x=2$. The authors' conclusion is that the observed change of the magnetic properties is attributed to a decrease of the strength of the negative $Mn^{2+}-Cr^{3+}$ superexchange interaction with increasing $Se$ concentration.

 We obtained, see figure 5, that the maximum of the magnetization is at $T^*$. Above $T^*$ the magnetization of the system is equal to the magnetization of sublattice $A$ spins. If we extrapolate this curve below $T^*$ down to zero temperature we will obtain a value close to $2s_1\mu_B$, where $s_1$ is the spin of the sublattice $A$ spin operators. The experimental figures \cite{HBMM3} show that extrapolations give one and the same result for all values of $x$. One can accept the fact that the $Se$ concentration do not influence over the value of  sublattice $A$ spin and $s_1=1.5$.

Below $T^*$ the magnetization is a sum of sublattice $A$ and $B$ magnetization. Hence, the magnetization at zero temperature is equal to $2(s_1-s_2)\mu_B$. Therefore, one can determine the sublattice $B$ spin $s_2$. The results of the theoretical calculations of magnetization, in Bohr magnetons, are depicted in figure 8 for parameters $s_1=1.5,\,J_1/J=0.47,\,J_2/J=0.001$ and $s_2=1$-curve \textbf{a}(black); $s_2=0.7$-curve \textbf{b} (red), and $s_2=0.4$-curve \textbf{c} (blue).
\begin{figure}[!ht]
\epsfxsize=9.cm \hskip -0.3cm \epsfbox{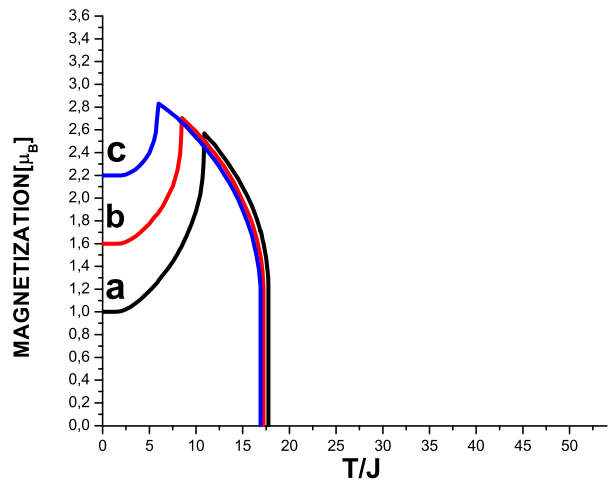} \caption{(color
online)\, The magnetization $2M^A+2M^B$ as function of $T/J$ for
$J_1/J=0.47$, $J_2/J\,=\,0.001$,\, $s_1=1.5$, and $s_2=1$-curve
\textbf{a}\,(black), $s_2=0.7$-curve \textbf{b}\,(red),
$s_2=0.4$-curve \textbf{c}\,(blue)}\label{fig8}
\end{figure}
The temperature and magnetization axis are chosen in accordance with experimental figure.
Comparing figure 94 in \cite{HBMM3} and figure 7 in the present paper, one concludes that the effective sublattice $B$ spin $s_2$ decreases with increasing Se concentration, and this is the origin of the anomalous temperature variation of magnetization. The figure 8 shows that the present calculations capture the essential  features of the system; increasing the $Se$ concentration (decreasing $s_2$) leads to a decrease of  $N\acute{e}el$ temperature, $T^*$ temperature decreases too, and the maximum of the magnetization  increases. Comparing the figure 8 in the present paper and figure 5 in \cite{Karchev08b} one realizes the importance of the present method of calculation for adequate reproducing the characteristic temperatures $T_N$, $T^*$, and the shape of the magnetization-temperature curves.

\subsection{\bf Vanadium spinel $MnV_2O_4$}

The spinel $MnV_2O_4$ is a two-sublattice ferrimagnet, with site $A$ occupied by the $Mn^{2+}$ ion, which is in the $3d^5$ high-spin configuration with quenched orbital angular momentum, which can be regarded as a simple $s=5/2$ spin. The B site is occupied by the $V^{3+}$ ion, which takes the $3d^{2}$ high-spin configuration in the triply degenerate $t_{2g}$ orbital and has orbital degrees of freedom. The measurements show that the setting in of the magnetic order is at $N\acute{e}el$ temperature $T_N=56.5K$ \cite{vanadium1} and that the magnetization has a maximum near $T^*=53.5K$. Below this temperature the magnetization sharply decreases and goes to zero when temperature approaches zero.

We consider a system which obtains its magnetic properties from $Mn$ and $V$ magnetic moments. Because of the strong spin-orbital interaction it is convenient to consider $jj$ coupling with $\textbf{J}^A=\textbf{S}^A$ and $\textbf{J}^B=\textbf{L}^B+\textbf{S}^B$. The sublattice $A$ total angular momentum is $j_A=s_A=5/2$, while the sublattice $B$ total angular momentum is $j_B=l_B+s_B$, with $l_B=3$, and $s_B=1$ \cite{vanadium1}.
Then the g-factor for the sublattice $A$ is $g_A=2$, and the atomic value of the $g_B$  is $g_B=\frac 54$. The sublattice $A$ magnetic order is antiparallel to the sublattice $B$ one and the saturated magnetization is $\sigma=2 \frac 52-\frac 54 4=0$, in agreement with the experimental finding that the magnetization goes to zero when the temperature approaches zero.
The Hamiltonian of the system is
\bea \label{MnV1}
 H & = & - \kappa_A\sum\limits_{\ll ij \gg _A } {{\bf J}^A_{i}
\cdot {\bf J}^A_{j}}\,-\,\kappa_B\sum\limits_{\ll ij \gg _B } {{\bf J}^B_{i}
\cdot {\bf J}^B_{j}}\nonumber \\
& & +\,\kappa \sum\limits_{\langle ij \rangle} {{\bf J}^A_{i}
\cdot {\bf J}^B_{j}}\eea
The first two terms  describe the ferromagnetic Heisenberg intra-sublattice
exchange $\kappa_A>0, \kappa_B>0$, while the third term describes the inter-sublattice exchange which is antiferromagnetic $\kappa>0$.
To proceed we use the Holstein-Primakoff representation of the total angular momentum vectors ${\bf J}^A_{j}(a^+_j,a_j)$ and ${\bf J}^B_{j}(b^+_j,\,b_j)$, where $a^+_j,\,a_j$
and $b^+_j,\,b_j$ are Bose fields, and repeat the calculations from sections II and III. The magnetization of the system $g_A\,M^A\,+\,g_B\,M^B$ as a function of the temperature is depicted  in figure 9 for parameters\, $\kappa_A/\kappa\,=\,0.45$\, and \,$\kappa_B/\kappa\,=\,0.001$. The parameters are chosen so that the calculations to reproduce the experimental value of the ratio $T_N/T^*$.
\begin{figure}[!ht]
\epsfxsize=8.5cm 
\epsfbox{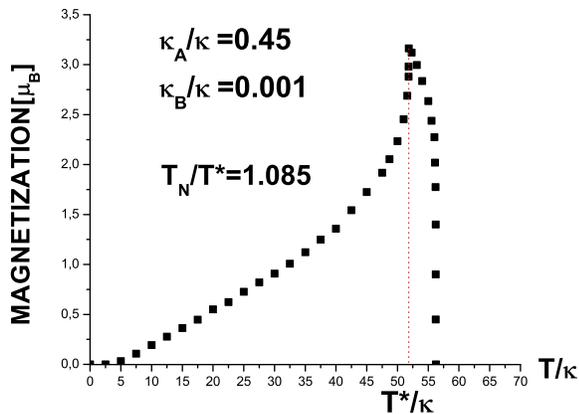} \caption{(color online)\,The magnetization
$g_A\,M^A\,+\,g_B\,M^B$ as a function of $T/\kappa$ for parameters\,
$\kappa_A/\kappa\,=\,0.45$\, and \,$\kappa_B/\kappa\,=\,0.001$.
}\label{fig9}
\end{figure}

The profile of the magnetization-temperature curve is in a very good agrement with the experimental zero-field cooling (ZFC) magnetization curves \cite{vanadium2,vanadium3}. The anomalous temperature dependence of the magnetization is reproduced, but there is an
important difference between the interpretation of the experimental results in \cite{vanadium1,vanadium2,vanadium3,vanadium4,vanadium5}, and the present theoretical results. In the experimental papers $T_N$ is the temperature at which both the $Mn$ and $V$ magnetization become equal to zero. The present theory predicts two phases: at low temperatures $(0,T^*)$ sublattice $Mn$ magnetization and sublattice $V$ magnetization contribute to the magnetization of the system, while at  high temperatures  $(T^*,T_N)$ only $Mn$ ions have non-zero spontaneous magnetization.
The vanadium sublattice magnetization set in at $T^*$, and evidence for this is the abrupt decrease of magnetization below $T^*$, which also indicates that the magnetic order of vanadium electrons is anti-parallel to the order of $Mn$ electrons.

For samples cooled in a field (FC magnetization) the field leads to formation of a single domain and, in addition, increases the chaotic order of the spontaneous magnetization of the vanadium sublattice, which is antiparallel to it. As a result the average value of the vanadium magnetic order decreases and does not compensate the $Mn$ magnetic order. The magnetization curves depend on the applied field, and do not go to zero. For a larger field the (FC) curve increases when temperature decreases below $N\acute{e}el$ temperature . It has a maximum at the same temperature $T^*<T_N$ as the ZFC magnetization, and a minimum at $T_1^*<T^*$. Below $T_1^*$  the magnetization increases monotonically when temperature approaches zero.

The experiments with samples cooled in field (FC magnetization) provide a new opportunity to clarify the magnetism of the manganese vanadium oxide spinel. The applied field is antiparallel with vanadium magnetic moment and strongly effect it. On the other hand, the experiments show that there is no difference between ZFC and FC magnetization curves when the temperature runs over the interval ($T^*,T_N$) \cite{vanadium2,vanadium3}. They begin to diverge when the temperature is below $T^*$. This is in accordance with the theoretical prediction that the vanadium magnetic moment does not contribute the magnetization when $T>T^*$ and $T^*$ is the temperature at which the vanadium ions start to contribute the magnetization of the system.
Because of the strong field, the two vanadium bands are split and the magnetic moment of one of the $t_{2g}$ electrons is reoriented to be parallel with the field and magnetic order of the $Mn$ electrons. The description of this case is more complicate and requires three magnetic orders to be involved. When $T^*<T<T_N$ only $Mn$ ions have non zero spontaneous magnetization. At $T^*$ vanadium magnetic order antiparallel to the magnetic order of $Mn$ sets in and partially compensates it. Below $T_1^*$ the reoriented electron gives contribution, which explains the increasing of the magnetization of the system when the temperature approaches zero. A series of experiments with different applied field could be decisive for the confirmation or rejection of the $T^*$ transition. Increasing the applied field one expects increasing of $T^*_1$ and when the field is strong enough, so that all vanadium electrons are reoriented, an anomalous increasing of magnetization below $T^{*}$ would be obtained as within the ferromagnetic
phase of $UGe_2$ \cite{2fmp6}.

\section{\bf Summary}

In summary, I have worked out a renormalized spin-wave theory and its extension to describe the two phases $(0,T^*)$ and $(T^*,T_N)$ of a two sublattice ferrimagnet. Comparing the figure 4 in the present paper and figure 4 in \cite{Karchev08b} and figure 8 in the present paper and figure 5 in \cite{Karchev08b} one becomes aware of the relevance of the present calculations for the accurate reproduction of the basic features of the system near the characteristic temperatures $T_N$ and $T^*$.

The present theory of ferrimagnetism permits to consider more complicate systems such as $CeCrSb_3$ compound \cite{Jackson1} or the spinel $Fe_3O_4$ which are two sublattice ferrimagnets but with three spins.

\section{Acknowledgments}

This work was partly supported by a Grant-in-Aid DO02-264/18.12.08 from NSF-Bulgaria.

\begin{appendix}
\section{}

To make more transparent the derivation of the equations for the Hartree-Fock parameters Eq.(\ref{rsw14}) I consider the first term (the sublattice $A$ term) in the Hamiltonian of the magnon-magnon interaction Eq.(\ref{rsw6}). To write this term in the Hartree-Fock  approximation one represents the product of two Bose operators in the form
\be\label{App1}
a^+_i a_j\,=\,a^+_i a_j\,-\,<a^+_i a_j>\,+\,<a^+_i a_j> \ee and neglects all terms $(a^+_i a_j\,-\,<a^+_i a_j>)^2$ in the four magnon interaction Hamiltonian. The result is
\bea\label{App2}
\frac 12 a^+_i a_j a^+_i a_i&\approx&-<a^+_i a_{j}><a^+_i a_i> \nonumber \\& + & <a^+_i a_{j}> a^+_i a_i+a^+_ia_{j}<a^+_i a_i> \nonumber \\
\frac 12 a^+_{j} a_i a^+_{j} a_{j} &\approx & -<a^+_{j} a_i><a^+_{j} a_{j}> \nonumber \\ & + & <a^+_{j} a_i> a^+_{j} a_{j}+a^+_{j} a_i <a^+_{j} a_{j}> \nonumber \\
\frac 12 a^+_{j} a_j a^+_i a_{j} &\approx & -<a^+_{j} a_{j}><a^+_i a_{j}>  \\ & + & <a^+_{j} a_{j}>a^+_i a_{j}+a^+_{j} a_{j}<a^+_r a_{j}> \nonumber\\
a^+_i a_i a^+_{j} a_{j} &\approx & -<a^+_i a_i>< a^+_{j} a_{j}> \nonumber \\ & + & <a^+_i a_i> a^+_{j} a_{j}+a^+_i a_i <a^+_{j} a_{j}> \nonumber \\
&-& <a^+_i a_{j}>< a^+_{j} a_i> \nonumber \\ & + &  <a^+ _i a_{j}> a^+_{j} a_i+a^+_{j} a_i <a^+_i a_{j}> \nonumber
\eea
The Hartree-Fock approximation of the sublattice $A$ part of the Hamiltonian of magnon-magnon interaction reads
\bea\label{App3}
& & \frac 14 J_1 \sum\limits_{\ll ij \gg _A }\left[a^+_i a^+_j( a_i-a_j)^2 + (a^+_i- a^+_j)^2  a_i a_j\right] \nonumber\\  & \approx & 12NJ_1 s_1^2 \left(u_1-1\right)^2 \\
& + & J_1 s_1 \left(u_1-1\right)\sum\limits_{\ll ij \gg _A }\left( a^+_i a_i\,+\,a^+_{j} a_{j}\,-\,a^+_{j} a_i\,-\,a^+_i a_{j}\right) \nonumber \eea
where the Hartree-Fock parameter $u_1$ is defined by the equation
\be\label{App4}
u_1\,=\,1\,-\,\frac {1}{6 s_1}\frac 1 N \sum\limits_{k\in B_r}e_k <a^+_k a_k> \ee
Combining the sublattice $A$ part of the Hamiltonian Eq.(\ref{rsw5}) (the first term) and  Eq.(\ref{App3}) one obtaines the Hartree-Fock approximation for the sublattice $A$ part of the Hamiltonian
\bea\label{App5}
H^A & \approx & 12NJ_1 s_1^2 \left(u_1-1\right)^2 \\
& + & J_1 s_1 u_1 \sum\limits_{\ll ij \gg _A }\left( a^+_i a_i\,+\,a^+_{j} a_{j}\,-\,a^+_{j} a_i\,-\,a^+_i a_{j}\right) \nonumber \eea

In the same way one obtains the Hartree-Fock approximation of the sublattice $B$ and inter sublattices parts of the Hamiltonian. The result is the $H_{HF}$ Hamiltonian Eqs.(\ref{rsw7},\ref{rsw8},\ref{rsw9}).

To calculate the thermal average $ <a^+_k a_k>$, in the Eq.(\ref{App4}), one utilizes the Hamiltonian $H_{HF}$. Therefor, the matrix element  depends on the Hartree-Fock parameters, and equation (\ref{App4}) is one of the self consistent equations for these parameters.

The matrix element can be represented in terms of $\alpha_k (\alpha_k^+)$ and $\beta_k (\beta_k^+)$ Eq.(\ref{rsw12a})
\be\label{App6}
<a^+_k a_k>\,=\,u_k^2 \,n_k^{\alpha}\, +\, v_k^2\, n_k^{\beta}\, +\, v_k^2 \ee
where $n_k^{\alpha}=<\alpha_k^+\alpha_k>,\,n_k^{\beta}=<\beta_k^+\beta_k>$ are the Bose functions of $\alpha$ and $\beta$ excitations.
  Substituting the thermal average in Eq.(\ref{App4}) with Eq.(\ref{App6}), one obtains that equation (\ref{App4}) is exactly the first equation of the system Eq.(\ref{rsw14}) which in turn is obtained from the first of the equations (\ref{rsw13}).

\end{appendix}

\end{document}